\documentclass[siam10.clo]{siamltex}

\begin{document}
\title{A fifth order expansion for the distribution function of the maximum likelihood estimator}

\author{Shanti Venetiaan\thanks{Institute for Graduate studies and research, Anton de Kom Universiteit van Suriname, Paramaribo, Suriname, s.venetiaan@uvs.edu} }

\maketitle

\begin{abstract}
In this paper, expansions for the maximum likelihood estimator of location and its distribution function are extended to fifth order. Since the proofs are straightforward extentions of proofs given in earlier papers for orders less than the fifth, they are not given here. The purpose of the paper is mainly to present the higher order expansions.
\end{abstract}

\begin{keywords}asymptotic expansion, maximum likelihood estimator, Cornish-Fisher expansions \end{keywords}

\begin{AMS}62E20\end{AMS}

\pagestyle{myheadings} \thispagestyle{plain} \markboth{SHANTI VENETIAAN}{{\small FIFTH ORDER EXPANSION FOR THE MLE AND ITS DISTRIBUTION}}

\section{Introduction}
This paper serves as an extention to fifth order of the expansions for  the maximum likelihood estimator of location, its distribution function and the quantiles of the distribution function. The same framework will be used as in Venetiaan(2010), which is mainly derived from Chibisov(1973), Chibisov and Van Zwet(1984), and Hall (1992). We assume that $X_1$,...,$X_n$ are i.i.d. with common density $f(\cdot - \theta)$, which is absolutely continuous. Furthermore the random variables have finite Fisher information for location, i.e. $I(f) = \int (f'/f)^2 f<\infty$. Our stronger regularity conditions, follow straightforwardly from the conditions in Venetiaan (2010). As a consequence, we will not give any proofs of the results, because these would be just adjustments of the detailed proofs given in earlier papers, mentioned above. In Section 2 the results are stated.

\subsection{Notation}
We will be using the following notation.
\begin{eqnarray} \eta_2 &= & E\psi_2^2(X_1)/I^2(f), \eta_3 = E\psi_1^3(X_1)/I^{3/2}(f), \eta_4=E\psi_1^4(X_1)/I^2(f),\nonumber\\
\eta_5&=& E\psi_1^5(X_1)/I^{5/2}(f), \eta_6=E(\psi_2(X_1)\psi_3(X_1))/I^{5/2}(f)\nonumber\\
\eta_7&=&E\psi_1^6(X_1)/I^3(f),\eta_8 = E\psi_2^3(X_1)/I^3(f),\eta_9 = E\psi_3^2(X_1)/I^3(f),\\
 \eta_{10} &=& E\psi_1\psi_2\psi_3(X_1)/I^3(f)\nonumber\\
 \quad& &\mbox{with} \quad\psi_i(x) = \frac{f^{(i)}}{f}(x).\nonumber\end{eqnarray}

\section{Results}
In this section we will state the results. These include a fifth order expansion for the maximum likelihood estimator of location, a fifth order expansion for its distribution function and a Cornish-Fisher expansion for its quantiles.

\vskip 4mm

Let $\hat \theta_n$ denote the maximum likelihood estimator of location, namely $\hat \theta_n$ satisfies
$$L_n(\hat \theta_n) =\inf_{\theta \in \mathbf R}L_n(\theta)$$
where $L_n(\theta) = n^{-1}\sum_{i=1}^n \rho(X_i - \theta)$, with
$\rho(\cdot) = -\log f(\cdot)$.

\vskip 8mm
The first result considers a fifth order expansion of the maximum likelihood estimator.

\begin{theorem}\label{stelling1} Let $X, X_1,...,X_n$ be i.i.d. with common density $f(\cdot - \theta_0)$.
Let $\hat \theta_n$ be the MLE and let $\rho(\cdot)$ satisfy the
following conditions.

(1)For all $K \subset \mathbf{R}$ compact, $\sup_{\theta\in K}
E_{\theta} \rho^2(X) = A < \infty$.

(2) $\rho(\cdot)$ is six times differentiable.

(3) There exists a finite function $R(\cdot)$ and a $\delta > 0$
such that, for every $y \in        \mathbf R$, $|\theta| < \delta
$:
\[ |\rho^{(6)}(y) -\rho^{(6)}(y-\theta)|\le R(y)|\theta|\quad\mbox{and}\quad E_0 R^{3}(X) < \infty\]

(4) $E_0 |\rho^{(\alpha)}(X)|^6 < \infty$ for $\alpha = 1,...,6$.

 Then $\hat\theta_n$ admits a stochastic expansion 

\begin{eqnarray}\sqrt n(\hat \theta_n -\theta_0) &=&  \frac{\xi_1}{a_2}
+\frac1{\sqrt{n}}(\frac{-\xi_1\xi_2}{a_2^2}+\frac{a_3\xi_1^2}{2a_2^3})+ \frac1{n}(\frac{\xi_1\xi_2^2}{a_2^3} - \frac{3a_3\xi_1^2\xi_2}{2a_2^4}+\frac{\xi_1^2\xi_3}{2a_2^3}+\frac{a_3^2\xi_1^3}{2a_2^5}-\frac{a_4\xi_1^3}{6a_2^4})\nonumber\\
&\quad + &\frac1{n\sqrt n}\biggl[\frac{3a_3\xi_1^2\xi_2^2}{a_2^5}+\frac{5a_3^3\xi_1^4}{8a_2^7}-\frac{5a_3a_4\xi_1^4}{12a_2^6}-\frac{3\xi_1^2\xi_2\xi_3}{2a_2^4}-\frac{5a_3^2\xi_1^3\xi_2}{2a_2^6}+\frac{a_5\xi_1^4}{24a_2^5}\label{eq:no4}\\
&\qquad +
&\frac{a_3\xi_1^3\xi_3}{a_2^5}+\frac{2a_4\xi_1^3\xi_2}{3a_2^5}-\frac{\xi_1^3\xi_4}{6a_2^4}-\frac{\xi_1\xi_2^3}{a_2^4}
\biggr]\nonumber\\
&\quad +& \frac1{n^2}\biggl[-\frac{5a_3\xi_1^2\xi_2^3}{a_2^6}-\frac{5a_4\xi_1^4\xi_3}{12a_2^6}+ \frac{15a_3^2\xi_1^3\xi_2^2}{2a_2^7}-\frac{35a_3^3\xi_1^4\xi_2}{8a_2^8}+\frac{5a_3a_4\xi_1^4\xi_2}{2a_2^7}-\frac{5a_5\xi_1^4\xi_2}{24a_2^6}\nonumber\\
&\quad-&\frac{7a_3^2a_4\xi_1^5}{8a_2^8}-\frac{a_6\xi_1^5}{120a_2^6}+ \frac{\xi_1\xi_2^4}{a_2^5}+ \frac{\xi_1^4\xi_5}{24a_2^5}+ \frac{a_4^2\xi_1^5}{12a_2^7}\nonumber\\
&\quad +& \frac{7a_3^4\xi_1^5}{8a_2^9} + \frac{\xi_1^3\xi_3^2}{2a_2^5}+ \frac{3\xi_1^2\xi_2^2\xi_3}{a_2^5}+ \frac{15a_3^2\xi_1^4\xi_3}{8a_2^7}- \frac{5a_3\xi_1^4\xi_4}{12a_2^6}\nonumber\\
&\quad -&\frac{5a_4\xi_1^3\xi_2^2}{3a_2^6}+ \frac{a_3a_5\xi_1^5}{8a_2^7}-\frac{5a_3\xi_1^3\xi_2\xi_3}{a_2^6}+ \frac{2\xi_1^3\xi_2\xi_4}{3a_2^5}\biggr]\nonumber\\
&\quad +&\gamma_n,\nonumber\end{eqnarray} 
where  $\xi_{j}$ denotes
the normalized sum of the independent random variables
$\rho^{(k)}(X_i)$.
\begin{equation}\xi_j = \frac1{\sqrt{n}}\sum_{i=1}^n (\rho^{(j)}(X_i)-a_j),\quad a_j = E_0\rho^{(j)}(X)\quad\mbox{for}\quad j=1,...,6.\end{equation}
Furthermore, for any sequence of positive constants
$\{\epsilon_n\}$, with $\epsilon_n\sqrt n(\log n)^{-2}\to \infty$,
we have
\begin{equation}P_0(|\gamma_n|\ge
\frac{\epsilon_n}{n^2})=o(\frac1{n^2}).\label{claim}\end{equation}
(end of Theorem \ref{stelling1})
\end{theorem}

\vskip 14mm
We now present the expansion for the distribution function.
\vskip 8mm
\begin{theorem}\label{stelling2}
Let the conditions of Theorem \ref{stelling1} hold, then
the distribution function of the MLE admits an
 Edgeworth expansion to order $n^{-2}$, namely

\begin{eqnarray}  G_n(x) &=& \Phi(x) -\frac{\eta_3 (x^2+2)}{12\sqrt n}\phi(x)\nonumber\\
&\quad +&\frac{1}{n}\biggl[-\frac{\eta_3^2}{288} x^5+(\frac18-\frac{\eta_2}6+\frac{5\eta_4}{72}+\frac{\eta_3^2}{72})x^3+(-\frac{\eta_4}{24}+\frac{\eta3^2}{24}+\frac18)x\biggr]\phi(x)\nonumber\\
&\quad+&\frac1{n\sqrt n}\biggl[-\frac{\eta_3^3}{10368}
x^8+(\frac{\eta_3}{96}-\frac{\eta_2\eta_3}{72}+\frac{19\eta_3^3}{10368}+\frac{5\eta_3\eta_4}{864})
x^6\label{dist}\nonumber\\
&\quad+&(-\frac{\eta_3\eta_4}{72}-\frac{\eta_5}{30}+\frac{19\eta_3^3}{1728}+\frac{\eta_6}8)x^4\nonumber\\
&\quad+&(\frac{35\eta_3^3}{864}+\frac{\eta_3}{32}+\frac{\eta_5}{80}-\frac{5\eta_3\eta_4}{96})x^2-
\frac{5\eta_3\eta_4}{48}+\frac{35\eta_3^3}{432}+\frac{\eta_3}{16}+\frac{\eta_5}{40}\biggr]\phi(x)\nonumber\\
&\quad+&\frac1{n^2}\biggl[  -\frac{\eta_3^4}{497664}x^{11}+(\frac{43}{497664}\eta_3^4-\frac{\eta_2\eta_3^2}{1728}+ \frac{\eta_3^2}{2304}+\frac{5\eta_3\eta_4^2}{20736})x^9\nonumber\\
&\quad+&(-\frac{5\eta_3^2}{1152}+ \frac{\eta_3^4}{3456}+ \frac{\eta_2\eta_3^2}{192}+ \frac{\eta_3\eta_6}{96}-\frac{\eta_3\eta_5}{360}-\frac{11\eta_3^2\eta_4}{3456}\nonumber\\
&\qquad-&\frac{\eta_2^2}{72}+\frac{\eta_2}{48}+\frac{5\eta_2\eta_4}{432}-\frac{1}{128}-\frac{5\eta_4}{576}-\frac{25\eta_4^2}{10368})x^7\nonumber\\
&\quad+& (-\frac{13\eta_2}{24}+ \frac{205\eta_4}{576} + \frac{\eta_9}{120}-\frac{\eta_8}{240}+ \frac{61\eta_{10}}{120}-\frac{731\eta_7}{3600}\nonumber\\
&\qquad +& \frac{23\eta_4^2}{3456}- \frac{\eta_2\eta_4}{72}+ \frac{287\eta_3^2}{2304}- \frac{5\eta_3^4}{6912}+ \frac{7}{384}-\frac{7\eta_3^2\eta_4}{768}\nonumber\\
&\qquad-& \frac{\eta_3\eta_6}{12}+ \frac{23\eta_3\eta_5}{960}+ \frac{\eta_2\eta_3^2}{48})x^5\nonumber\\
&\quad +&(\frac{23}{48}\eta_2-\frac{181\eta_4}{576}+ \frac{\eta_8}{24}-\frac{5\eta_{10}}{12}+ \frac{53\eta_7}{360}+ \frac{7\eta_4^2}{1152}\nonumber\\
&\qquad-& \frac{\eta_2\eta_4}{48}-\frac{77\eta_3^2}{576}- \frac{35\eta_3^4}{3456}+ \frac{5\eta_3^2\eta_4}{1728}- \frac{\eta_3\eta_6}{6}\nonumber\\
&\qquad+& \frac{\eta_3\eta_5}{24}+ \frac{5\eta_2\eta_3^2}{144}+\frac{5}{384})x^3\nonumber\\
&\quad +& (\frac{\eta_4}{64}+ \frac{\eta_7}{240}- \frac{5\eta_4^2}{384}- \frac{\eta_3^2}{64}- \frac{35\eta_3^4}{1152}+ \frac{1}{128}+ \frac{35\eta_3^2\eta_4}{576}-\frac{\eta_3\eta_5}{48})x  \biggr]\phi(x)\nonumber\\
&+& o(\frac1{n^2})\label{exGn}
\end{eqnarray}
\end{theorem}


\vskip 0.8cm

The last result is about a Cornish-Fisher expansion for the inverse of the distribution function of the MLE.

\vskip 8mm

\begin{theorem}\label{stelling3}

If the conditions of Theorem \ref{stelling1} are fulfilled, then the inverse of the distribution function of the MLE admits a Cornish-Fisher expansion.

\begin{eqnarray}  G_n^{-1}(v) &=& \Phi^{-1}(v)+(\frac{\eta_3}{12\sqrt n}(\Phi^{-1}(v))^2+2)\nonumber\\
&+& \frac1{n}\biggl[(-\frac{\eta_3^2}{72}-\frac{5\eta_4}{72}+ \frac{\eta_2}6-\frac18)(\Phi^{-1}(v))^3
+ (-\frac{\eta_3^2}{36}+ \frac{\eta_4}{24}-\frac18)\Phi^{-1}(v)\biggr]\nonumber\\
&+& \frac1{n\sqrt n}\biggl[(-\frac{\eta_3\eta_4}{144}+\frac{\eta_2\eta_3}{24}-\frac{\eta_3}{48}-\frac{19\eta_3^3}{1728}-\frac{\eta_6}{8}+\frac{\eta_5}{30})(\Phi^{-1}(v))^4\nonumber\\
&+&(\frac{\eta_3\eta_4}{48}+ \frac{\eta_2\eta_3}{12}-\frac{5\eta_3}{48}-\frac{67\eta_3^3}{1296}-\frac{\eta_5}{80})(\Phi^{-1}(v))^2-\frac{113\eta_3^3}{1296}-\frac{\eta_5}{40}+\frac{\eta_3\eta_4}{9}-\frac{\eta_3}{12}\biggr]\nonumber\\
&+& \frac1{n^2}\biggl[(\frac{7\eta_2}{16}-\frac{59\eta_4}{192}-\frac{23\eta_3^2}{192}-\frac{61\eta_{10}}{120}-\frac{\eta_2\eta_4}{16}+ \frac{731\eta_7}{3600}\nonumber\\
&+& \frac{\eta_8}{240}-\frac{\eta_9}{120}+ \frac{19\eta_3^2\eta_4}{1728}-\frac{7\eta_2\eta_3^2}{288}-\frac{17\eta_3\eta_5}{1440}+ \frac{\eta_3\eta_6}{24}\nonumber\\
&+& \frac{\eta_2^2}{12}+ \frac{37\eta_4^2}{3456}+ \frac{\eta_3^4}{1728}+ \frac{5}{384})(\Phi^{-1}(v))^5\nonumber\\
&+&(\frac{1}{24}- \frac{9\eta_2}{16}+ \frac{\eta_4}{3}+ \frac{85\eta_3^2}{576}+ \frac{5\eta_{10}}{12}+\frac{7\eta_2\eta_4}{144}-\frac{53\eta_7}{360}\nonumber\\
&-& \frac{\eta_8}{24}+ \frac{11\eta_3^2\eta_4}{1728}-\frac{\eta_2\eta_3^2}{24}-\frac{7\eta_3\eta_5}{360}+ \frac{\eta_3\eta_6}{12}-\frac{\eta_4^2}{54}\nonumber\\
&+& \frac{19\eta_3^4}{3888})(\Phi^{-1}(v))^3+(\frac1{128}-\frac{5\eta_4}{192}+ \frac{\eta_3^2}{288}-\frac{\eta_7}{240}-\frac{5\eta_3^2\eta_4}{96}\nonumber\\
&+&\frac{\eta_2\eta_3^2}{72}+ \frac{\eta_3\eta_5}{60}+ \frac{17\eta_4^2}{1152}+ \frac{65\eta_3^4}{3888})\Phi^{-1}(v)\biggr]\label{eq:no3}
\end{eqnarray}

\end{theorem} 


\enddocument